\documentclass[letter]{aa}  

\usepackage{graphicx}
\usepackage{txfonts}
\usepackage[colorlinks=True,citecolor=blue,linkcolor=blue,urlcolor=blue]{hyperref}
\usepackage{multirow}
\usepackage{orcidlink}

\newcommand{\hi}{H} 
\newcommand{\hei}{He} 
\newcommand{\pn}{SMP~LMC~58}
\newcommand{\kmps}{km s$^{-1}$}

\newcommand{\ajs}[1]{{{#1}}}
\newcommand{\refedit}[1]{{{#1}}}

\begin{document} 

   \title{An accurate measurement of the  spectral resolution of the JWST Near Infrared Spectrograph}

   \titlerunning{Accurate resolution of the JWST NIRSpec}
   \authorrunning{Shajib et al.}

   \subtitle{}

   \author{Anowar~J.~Shajib\orcidlink{0000-0002-5558-888X},\inst{ \ref{uchicago},\ref{kicp},\ref{cassa}}
		Tommaso~Treu\orcidlink{0000-0002-8460-0390},\inst{\ref{ucla}}
          Alejandra Melo\orcidlink{0000-0002-6449-3970},\inst{\ref{mpa}, \ref{tum}}
          Guido Roberts-Borsani\orcidlink{0000-0002-4140-1367},\inst{\ref{ucl}}
          Shawn~Knabel\orcidlink{0000-0001-5110-6241},\inst{\ref{ucla}}
		Michele~Cappellari\orcidlink{0000-0002-1283-8420},\inst{\ref{oxford}}
         and
    Joshua~A.~Frieman\orcidlink{0000-0003-4079-3263}\inst{\ref{uchicago},\ref{kicp},\ref{slac}}
          }
    
    \institute{\footnotesize
    Department  of  Astronomy  \&  Astrophysics,  University  of Chicago, Chicago, IL 60637, USA; \email{ajshajib@uchicago.edu}\label{uchicago}
    \and
    Kavli Institute for Cosmological Physics, University of Chicago, Chicago, IL 60637, USA \label{kicp}
    \and
    Center for Astronomy, Space Science and Astrophysics, Independent University, Bangladesh, Dhaka 1229, Bangladesh \label{cassa}
    \and
    Department of Physics and Astronomy, University of California, Los Angeles, CA 90095, USA \label{ucla}
    \and
    Max Planck Institute for Astrophysics, Karl-Schwarzschild-Str.~1, Garching, 85748, Germany \label{mpa}
    \and
    Technical University of Munich, TUM School of Natural Sciences, Physics Department, James-Franck-Stra{\ss}e 1, 85748 Garching, Germany\label{tum}
    \and
    Department of Physics \& Astronomy, University College London, London, WC1E 6BT, UK \label{ucl}
    \and
    Sub-Department of Astrophysics, Department of Physics, University of Oxford, Denys Wilkinson Building, Keble Road, Oxford, OX1 3RH, UK \label{oxford}
    \and
    SLAC National Laboratory, 2575 Sand Hill Rd, Menlo Park, CA 94025 \label{slac}
    }
    
   \date{Received xxx, xxxx; accepted xxx, xxxx}

 
  \abstract{
   The spectral resolution ($R \equiv \lambda / \Delta \lambda$) of spectroscopic data is crucial information for accurate kinematic measurements. In this letter, we present a robust measurement of the spectral resolution of the JWST's Near Infrared Spectrograph (NIRSpec) \refedit{in fixed slit (FS) and integral field spectroscopy (IFS) modes. Due to the similarity of the utilized slit dimension if the FS mode to that of the shutters in the multi-object spectroscopy (MOS) mode, our resolution measurements in the FS mode can also be used for the MOS mode in principle.} We modeled \hi\ and \hei\ lines of the planetary nebula \pn\ using a Gaussian line spread function (LSF) to estimate the wavelength-dependent resolution for multiple disperser and filter combinations. We corrected for the intrinsic width of the planetary nebula's \hi\ and \hei\ lines due to its expansion velocity by measuring it from a higher-resolution X-shooter spectrum. We find that NIRSpec's in-flight spectral resolutions exceed the pre-launch estimates provided in the JWST User Documentation \refedit{by 11--53\% in the FS mode and by 1--24\% in the IFS mode across the covered wavelengths.} We recover the expected trend that the resolution increases with the wavelength within a configuration. The robust and accurate LSFs presented in this letter will enable high-accuracy kinematic measurements using NIRSpec for applications in cosmology and galaxy evolution.}


   \maketitle
%

\section{Introduction}

The JWST's Near Infrared Spectrograph \citep[NIRSpec;][]{Boker23} has enabled IR spectroscopy with an unprecedented combination of redshift reach, signal-to-noise ratio, and spatial resolution. One of its key applications is the study of stellar or gas kinematics in distant galaxies \citep[e.g.,][]{deGraaff24, Xu24a, Newman25}, helping us to understand their formation and evolution, the properties of gas outflows, and the impact of black hole and baryonic feedback \citep{DEugenio24}. Furthermore, accurate measurement of the stellar velocity dispersion of lensing galaxies also enables high-precision measurement of the Hubble constant and other cosmological parameters through the method of time-delay cosmography \citep{Shajib25, TDCOSMO25}.
	
Accurate measurement of the line spread function (LSF) or the spectral resolution \refedit{$R \equiv \Delta \lambda / \lambda$, where $\Delta \lambda$ is the full width at half maximum (FWHM),} is a prerequisite to robustly measuring the velocity dispersion $\sigma_{\star}$. The LSF induces additional broadening in the observed lines, which can be parametrized with $\sigma_{\rm inst}$ assuming a Gaussian LSF. The accuracy requirement on $\sigma_{\rm inst}$ to achieve a desired level of accuracy on $\sigma_\star$ can be estimated using the relation \citep{Knabel25}
   \begin{equation}
      \frac{\delta \sigma_{\star}}{\sigma_{\star}} \approx \frac{\delta \sigma_{\mathrm{inst}}}{\sigma_{\mathrm{inst}}}      \left( \frac{\sigma_{\mathrm{inst}}}{\sigma_{\star}} \right)^2.
   \end{equation}
Therefore, achieving a 1\% accuracy in the velocity dispersion, as desirable for precision cosmology \citep{Knabel25}, requires an estimate of NIRSpec's medium-resolution $\sigma_{\rm inst}$ accurate to 1.3\% and 5.5\% for $\sigma_{\star} = 150$ \kmps\ and 300 \kmps, respectively, for example. 
	
NIRSpec's LSF in the fixed-slit (FS) mode was previously measured from the narrow emission lines of the planetary nebula (PN) \pn\ \citep{Isobe23}. These authors find that the resolution is higher by $\sim$10--20\% than the pre-launch estimates provided by the JWST User Documentation (JDox).\footnote{\href{https://jwst-docs.stsci.edu/jwst-near-infrared-spectrograph/nirspec-instrumentation/nirspec-dispersers-and-filters}{JDox webpage for NIRspec dispersers and filters}} 
    These authors assumed that the intrinsic expansion velocity of the PN is negligible. However, if the true expansion velocity is near the upper limit of typical values \citep[i.e, $\lesssim$30 \kmps;][]{Jacob13}, its impact may not be non-negligible, especially for the high-resolution grating ($R \sim 2700$), which would require a $\sim$19\% correction ($\sim$2.7\% for medium resolution with $R \sim 1000$).
    \citet{Nidever24} find the resolution in the multi-object spectroscopy (MOS) mode's G140H/F100LP configuration to be $\sim$50--85\% higher than the JDox estimates, using narrow absorption lines in red giant stars.
    Through modeling with simulated data in the MOS mode, \citet{deGraaff24} attribute the discrepancy between direct measurements and JDox estimates to differences in the source morphology. These authors demonstrate that a point source would lead to $\sim$50-90\% higher resolution in the MOS G395H/F290LP configuration than the JDox estimates, which are based on uniformly illuminated slits. However, the predicted resolution by \citet{deGraaff24} did not account for the additional broadening introduced by the data reduction procedure. Hence, this value is not directly applicable to the reduced data, and a direct measurement from real data remains indispensable. 
		
In this letter, we provide a wavelength-dependent parametrization of NIRSpec's resolution for multiple disperser--filter combinations for all of its FS, integral field spectroscopy (IFS), and MOS modes. We measured the resolutions by fitting prominent \hi\ and \hei\ emission lines for the PN \pn\ using datasets from JWST calibration programs, which were specifically obtained for wavelength and LSF calibrations. \citet{Isobe23} also used a subset of this dataset for their LSF measurements. The same PN was also targeted by \citet{Jones23} to measure the LSF for JWST's Mid-Infrared Instrument (MIRI). Most of the fitted \hi\ lines have nearby weaker \hei\ multiplets, which are partially or fully blended and need to be accounted for to obtain an accurate characterization of the LSF. As a significant improvement over previous studies, we account for all the blended lines. We simultaneously fit the medium- and high-resolution spectra, thereby allowing lines that are deblended in the high-resolution spectrum to inform their relative strengths when fitting the medium-resolution one. Furthermore, we corrected for the PN's expansion velocity after directly measuring it from a higher-resolution ($R \sim 6500$) spectrum of the same PN obtained using the Very Large Telescope's (VLT) X-shooter instrument \citep{Vernet11}.

This letter is organized as follows. In Section \ref{sec:data}, we describe the JWST and X-shooter datasets we used. We describe our measurement method and provide the mesured values in Section \ref{sec:method}, and then conclude the letter in Section \ref{sec:conclusion}. Code scripts and notebooks used in this analysis are publicly available on GitHub.\footnote{ \url{https://github.com/ajshajib/nirspec_resolution}}.




\section{Datasets} \label{sec:data}


We describe the NIRSpec dataset used to measure the spectral resolutions in Section \ref{sec:nirspec_data} and the VLT X-Shooter spectra used to measure the PN's expansion velocity in Section \ref{sec:xshooter}.

\subsection{NIRSpec spectra} \label{sec:nirspec_data}

To provide instrument calibrations for the NIRSpec, the planetary nebula \pn\ (also known as IRAS 05248$-$7007) was observed with the calibration program CAL-1492 (PI: T.~Beck). We utilized the FS and IFS mode spectra from this program.\footnote{{We do not utilize the MOS-mode observations from the program CAL-1492, as the exposures were stepped at 25 mas steps across the full pitch of an MSA shutter. Thus, the LSF in the stacked data is expected to have broadened beyond that from a point source centered on the slit with only sub-pixel dithers applied. However, due to the similarity of the S200A1 slit dimension to the MSA shutters, the FS-mode values are theoretically expected to be a reasonable proxy for MOS-mode observations. Suitable MOS-mode datasets in the future can be easily analyzed with the code suite and notebooks we provided.}} \refedit{The FS mode used the S200A1 slit with 0\farcs2 width, while including two primary dither positions with spectral sub-pixel dithers to
improve the sampling. The IFS observations included a 4-point nod to improve sampling and enable background subtraction.} We obtained the reduced Level-3 spectra from the Mikulski Archive for Space Telescopes (MAST) archive, which had been reduced with the default JWST pipeline v1.18.0 \citep{Bushouse23} with CRDS context \texttt{jwst\_1364.pmap}. \refedit{Thus, the reduced and stacked data followed the standard steps in the pipeline.}

The FS-mode data were already extracted in the form of 1D spectra. For the IFS mode, we summed the spaxels within a circular aperture of radius 0\farcs55, centered on the PN, to obtain the corresponding 1D spectra. 

\subsection{VLT X-shooter specrum} \label{sec:xshooter}

The VLT X-shooter spectrum of the PN was obtained as part of an \textit{Euclid}-preparation program \citep[program ID 110.23Q7.001;][]{EuclidCollaboration23}. The spectrum was obtained with the VIS arm (covering 550--1020 nm) using a 1\farcs2 slit, which has a nominal resolution of $R \sim 6500$.\footnote{\url{https://www.eso.org/sci/facilities/paranal/instruments/xshooter/inst.html}} We obtained the observed raw dataset from the European Southern Observatory (ESO) Science Portal,\footnote{\url{https://archive.eso.org/scienceportal/}} along with the calibration files. We reduced the dataset to produce an extracted 1D spectrum using ESO's X-shooter data reduction pipeline v3.8.1 \citep{Modigliani10} with \textsc{EsoReflex} v2.11.5.

\section{Measurement of the NIRSpec resolution} \label{sec:method}

We adopted the Gaussian profile to characterize the NIRSpec LSF. By testing on a single emission line (Pa$\alpha$) that is not blended with other lines in both high and medium resolutions, we find that the LSF can be well-fitted with either a Gaussian or a Voigt profile, with the Voigt profile providing a better fit due to one additional degree of freedom. The Lorentzian profile, however, yields a significantly worse fit. 
We chose the Gaussian profile also because it is ubiquitously used to account for instrumental resolution in measurements of kinematics, for example, with the software program \textsc{pPXF} \citep{Cappellari17, Cappellari23}. The resolution parameter $R \equiv \lambda / \Delta \lambda$ relates to the associated instrumental dispersion $\sigma_{\rm inst}$ as $R = c / (2.355\sigma_{\rm inst})$, where $c$ is the speed of light. We parametrize the wavelength dependence of $\sigma_{\rm inst}(\lambda) = [{\sigma_{\rm inst}^\prime(\lambda)}^2 - \sigma_{\rm PN}^2]^{1/2} $ as
\begin{equation}\label{eq:parametrization}
    \sigma_{\rm inst} (\lambda) = \left[
    \left\{\frac{\sigma^{\prime}_{\rm piv}}{1 + \alpha\, (\lambda - \lambda_{\rm piv}) \, / \, [1\ \mu\rm{m}]} \right\}^2 - \sigma_{\rm PN}^2\right]^{1/2}, 
\end{equation}
where $\sigma_{\rm inst}^\prime(\lambda)$ is the instrumental dispersion not corrected for the PN's expansion velocity $\sigma_{\rm PN}$, $\sigma_{\rm piv}^\prime$ is the uncorrected instrumental dispersion at the pivot wavelength $\lambda_{\rm piv}$, and $\alpha$ is the coefficient in the inverse-linear dependence on wavelength. We chose the inverse-linear $\lambda$-dependence for $\sigma_{\rm inst}^\prime$ because the resolution $R  \propto 1/\sigma_{\rm inst}$ is found approximately linear in $\lambda$ from the pre-launch estimates, given that $\Delta\lambda$ tends to be approximately constant in wavelength units. We describe our measurement process for $\sigma_{\rm inst}^\prime (\lambda)$ in Section \ref{sec:jwst_fitting} and for $\sigma_{\rm PN}$ in Section \ref{sec:xshooter_fitting}.

\subsection{Emission line fitting from the NIRSpec spectra} \label{sec:jwst_fitting}

\begin{figure} [!t]
	\centering
\includegraphics[width=0.5\textwidth]{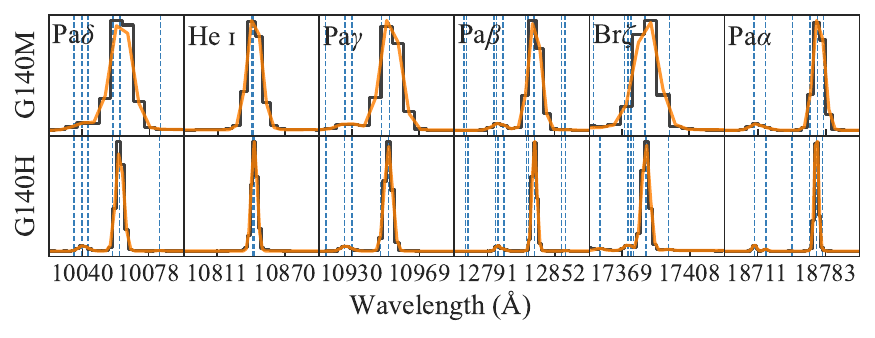}	\caption{\label{fig:nirspec_line_fits}
	Post-pixelized Gaussian profile fits (orange) to prominent emission-line groups (black) in the NIRSpec IFS spectra of the PN \pn. The IFS-mode G140 disperser is illustrated here as an example among the nine combinations analyzed in this letter. The top row corresponds to the G140M/F100LP configuration (i.e., medium resolution), and the bottom row to G140H/F100LP (i.e., high resolution). The principal line within each group is annotated in the corresponding panel. The individual line wavelengths fitted in each group are marked with dashed blue lines. 
    All of these lines are fitted simultaneously to directly constrain the parameters in $\sigma_{\rm inst}^\prime(\lambda)$ from Eq.~(\ref{eq:parametrization}).
	}
\end{figure}

We fit the spectra in wavelength slices around several prominent \hi\ and \hei\ emission lines. The details of the fitted wavelength ranges and the lists of lines within them are provided in Appendix \ref{app:line_list}. Prominent \hi\ lines are often accompanied by \hei\ lines in proximity due to the resemblance in their atomic structures. Furthermore, \hei\ lines are usually multiplets. These lines are partially or fully blended, especially in the medium resolution. For a given combination of mode, disperser, and filter, we simultaneously fit all the lines within the chosen wavelength slices both in the high and medium resolution spectra, while individually constraining $\sigma_{\rm piv}^\prime$ and $\alpha$ for both gratings. The emission lines were fitted with post-pixelized \refedit{(i.e., integrated within a pixel)} Gaussian LSFs. \ajs{The advantage of fitting all the lines together is that the information from all of them was aggregated to alleviate the issue that the LSF in most configurations is not critically sampled according to the Nyquist sampling theorem.} The advantage of fitting both gratings together is that the relative line strengths could first be robustly constrained from the high-resolution spectrum, where they are less blended, which were then fixed for fitting the blended lines in the medium resolution (see Figure \ref{fig:nirspec_line_fits}).
We also fit for the PN's relative motion $v_{\rm PN}$ \citep[previously measured as $295 \pm 23$  \kmps;][]{Reid06} without any dependency on the wavelength. We allowed the PN velocity to be independent between the medium- and high-resolution spectra to account for any residual offset in the wavelength calibration. In each wavelength slice, we fit the continuum level with a linear function. \refedit{We additionally allowed a Gaussian intrinsic scatter $\delta_{\rm intr}$ in the $\sigma_{\rm inst}$ fitted to each line group around the mean relation of Eq. \eqref{eq:parametrization}. Furthermore, we accounted for potential underestimation in the pipeline-produced uncertainties by allowing a systematic uncertainty floor $\varepsilon_{\rm syst}$ to vary freely for each resolution.} 
\refedit{Thus, we have 10 non-linear parameters characterizing the global properties of our generative model: $\sigma_{\rm piv}^\prime$, $\alpha$, $v_{\rm PN}$, $\delta_{\rm intr}$, and $\varepsilon_{\rm syst}$} for each resolution. The amplitudes of the line profiles and the continuum levels were solved through linear inversion. 
We obtained the posteriors of the non-linear parameters (provided in Table \ref{tab:param_values}) using the \textsc{nautilus} sampler \citep{Lange23}.

\begin{figure*} [!t]
	\centering
	\includegraphics[width=0.85\textwidth]{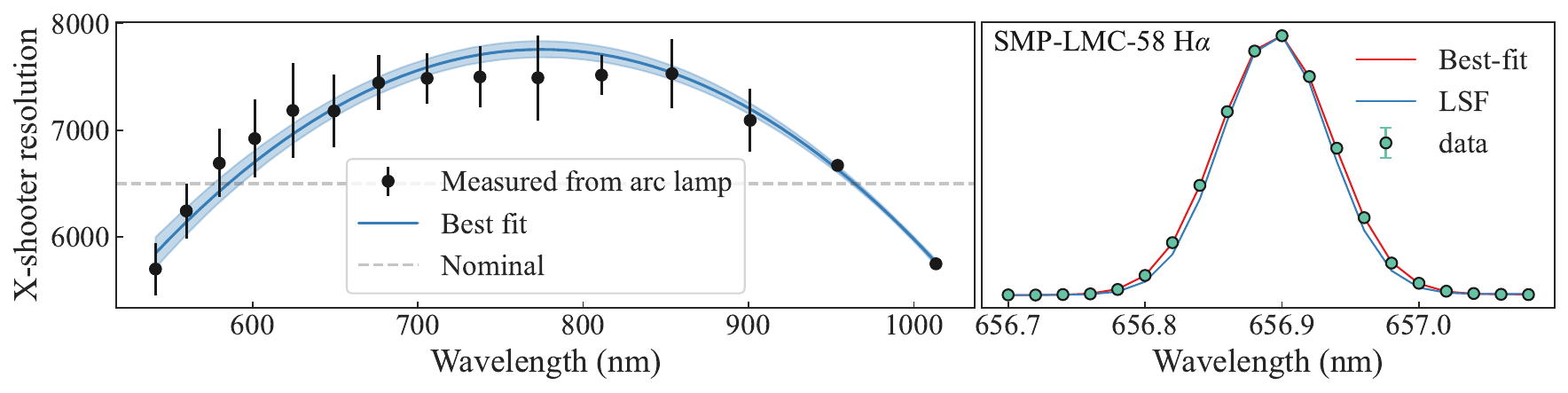}
	\caption{\label{fig:xshooter_lsf}
	Measurement of the PN's expansion velocity. \textit{Left-hand panel:} 
    Measurement of the X-shooter resolution (black points with error bars) from arc-lamp lines with the wavelength dependence modeled with a quadratic function (blue line, with the shaded region showing 1$\sigma$ uncertainty). \textit{Right-hand panel:} The H$\alpha$ line of the PN in the X-shooter spectra (emerald points). The error bars are too small to be noticeable here. The best-fit post-pixelized Gaussian profile is shown in red, while the width of the X-shooter LSF, as determined by the best fit in the left-hand panel, is shown in blue for comparison. Both sets of illustrated datapoints are simultaneously fit to infer the PN's expansion velocity $\sigma_{\rm PN} = 6.90 \pm 0.49$ \kmps.
	}
\end{figure*}

\subsection{Measurement of the PN expansion velocity} \label{sec:xshooter_fitting}

To measure the expansion velocity of the PN from the X-shooter spectrum, we require a precise measurement of the spectral resolution of the X-shooter itself. We used the arc-lamp spectrum that was taken for wavelength calibration to obtain a wavelength-dependent estimate of X-shooter's spectral resolution. We used the \textsc{PypeIt} software program \citep{Prochaska20} to measure the FWHMs of the arc-lamp lines in 15 wavelength bins within the covered wavelength range of 550--1020 nm (Figure \ref{fig:xshooter_lsf}). The uncertainty of the binned measurement was derived from the scatter in the individual-line measurements within each bin. The extent of the variation in the measured resolutions across the wavelength is qualitatively similar to that measured by \citet{Gonneau20}.  We fit a quadratic function $R_{\rm xsh}(\lambda)$ to the measured points. We also simultaneously fit a post-pixelized Gaussian function to the H$\alpha$ line, which is the most prominent H line within the covered wavelengths. Although a few \hei\ lines also fall within the covered wavelength range, they do not significantly add to the constraining power due to their line strengths being smaller by 1.5 orders of magnitude or more. We parametrized the H$\alpha$ line's Gaussian profile with standard deviation $[ \sigma_{\rm PN}^2 + c^2/\{2.355R_{\rm xsh}(\lambda_{\rm H\alpha})\}^2]^{-1/2}$, while allowing the peak position to vary to account for the PN's relative motion. We constrain the PN's expansion velocity $\sigma_{\rm PN} = 6.90 \pm 0.49$ \kmps.

\begin{figure}
	\centering
\includegraphics[width=1\columnwidth]{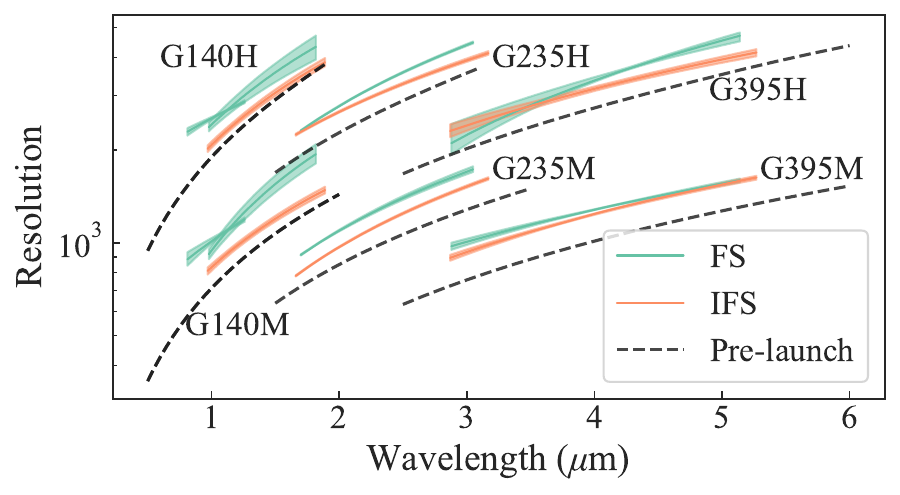}
\caption{\label{fig:nirspec_resolution}
	Comparison of our measured resolution curves in the FS (emerald) and IFS (orange) modes with the pre-launch estimates from JDox (dashed grey lines). The shaded region around each resolution curve signifies the 1$\sigma$ credible region. The in-flight resolutions are 11--53\% higher than the pre-launch estimates in the FS mode, and 1--24\% higher in the IFS mode (across the covered wavelengths).
	}
\end{figure}

We illustrate the measured resolution curves for all the mode--disperser combinations in Figure \ref{fig:nirspec_resolution} and compare them to the corresponding pre-launch estimates provided by JDox. The resolutions in the IFS mode \refedit{across the covered wavelengths} are higher by $\sim$1--24\% than the pre-launch estimates, whereas the resolutions in the FS mode are $\sim$11--53\% higher.

\begin{table}
\caption{ \label{tab:param_values}
Best-fit values for parameters in the expression of $\sigma_{\rm inst}^\prime(\lambda)$ in Eq.~(\ref{eq:parametrization}).}
\renewcommand{\arraystretch}{1.1}
\centering
\resizebox{\columnwidth}{!}{\begin{tabular}{llcccc}
\hline \hline
Disperser & Filter & $\lambda_{\rm piv}$ ($\AA$) & $\sigma^\prime_{\rm piv}$ (km s$^{-1}$) & $\alpha$ \\
\hline
\textbf{FS mode} &  &  &  &  \\ 
G140M & F070LP & 10111.02 & $124.84\pm2.37$ & $0.676\pm0.147$ \\ 
G140H & F070LP & 10111.02 & $\phantom{0}50.44\pm0.44$ & $0.500\pm0.083$ \\ 
G140M & F100LP & 13347.24 & $\phantom{0}94.99\pm5.10$ & $0.899\pm0.054$ \\ 
G140H & F100LP & 13347.24 & $\phantom{0}40.63\pm2.25$ & $0.703\pm0.082$ \\ 
G235M & F170LP & 22765.89 & $101.07\pm1.12$ & $0.478\pm0.018$ \\ 
G235H & F170LP & 22765.89 & $\phantom{0}40.20\pm0.12$ & $0.479\pm0.005$ \\ 
G395M & F290LP & 38474.54 & $102.95\pm0.52$ & $0.220\pm0.018$ \\ 
G395H & F290LP & 38474.54 & $\phantom{0}40.48\pm1.04$ & $0.340\pm0.061$ \\ 
\textbf{IFS mode} &  &  &  &  \\ 
G140M & F100LP & 13539.59 & $117.01\pm2.25$ & $0.668\pm0.035$ \\ 
G140H & F100LP & 13539.59 & $\phantom{0}46.36\pm1.05$ & $0.699\pm0.036$ \\ 
G235M & F170LP & 22938.52 & $112.74\pm0.59$ & $0.488\pm0.008$ \\ 
G235H & F170LP & 22938.52 & $\phantom{0}42.75\pm0.28$ & $0.398\pm0.010$ \\ 
G395M & F290LP & 38891.71 & $105.74\pm0.56$ & $0.251\pm0.014$ \\ 
G395H & F290LP & 38891.71 & $\phantom{0}41.98\pm0.70$ & $0.239\pm0.026$ \\ 
\hline
\end{tabular}
}
\tablefoot{
The value of $\sigma_{\rm PN}$ in Eq.~(\ref{eq:parametrization}) is measured in Section \ref{sec:xshooter_fitting} to be $\sigma_{\rm PN} = 6.90 \pm 0.49$ \kmps. 
}
\end{table}

\section{Conclusion} \label{sec:conclusion}

In this letter, we measured NIRSpec's spectral resolution in multiple configurations for in \refedit{the FS and IFS modes. Due to the similarity in slit dimensions, the FS-mode resolution can be used as a reasonable proxy for the MOS mode.} We provide a wavelength-dependent parametrization of the spectral resolution in terms of the instrumental dispersion $\sigma_{\rm inst}$ by fitting multiple \hi\ and \hei\ emission lines across the observed wavelength range in the PN \pn. We incorporated a correction for the intrinsic width of the PN's lines, which we measured using a higher-resolution X-shooter spectrum of the same PN. \refedit{Our measured resolutions are directly applicable for point source spectra; however, caution is advised when using them for extended sources.}

The measured resolutions increase with wavelength in all configurations, as expected from the pre-launch estimates provided by JDox. For the FS mode, our measured resolution values extend up to 20\% higher than those found by \citet{Isobe23}, potentially due to our methodology allowing multiple lines to be fitted within a blended line group. \refedit{Our measured resolution in the G140H/F100LP configuration is smaller than that found by \citet{Nidever24}, which may be a difference between single-exposure and stacked spectra. Therefore, caution is advised when applying our measured value for datasets with drastically different dithering procedures from CAL-1492.}

The parametric wavelength-dependent functions we provide will enable accurate measurements of kinematics from the NIRSpec data, as performed, e.g., by \citet{TDCOSMO25} and \citet{Shajib25}. Although we do not provide the MOS-mode values at the medium resolution, the FS-mode values in medium resolution can be used as a substitute following \citet{Isobe23}, as the slit width for the analyzed data in the FS mode is the same as that of the multi-shutter-array slitlets.

\begin{acknowledgements}
We thank David Nidever for helpful comments as the referee, which improved this manuscript. We thank Hsiao-Wen Chen, Danial Rangavar Langeroodi, Mark Morris, and Stefan Noll for useful discussions.
This work is based on observations made with the NASA/ESA/CSA James Webb Space Telescope. The data were obtained from the Mikulski Archive for Space Telescopes at the Space Telescope Science Institute, which is operated by the Association of Universities for Research in Astronomy, Inc., under NASA contract NAS 5-03127 for JWST. These observations are associated with programs \#1125 and 1492. The specific observations analyzed can be accessed via \url{https://dx.doi.org/10.17909/m3c0-3t58}.
AJS and TT acknowledge support from NASA through STScI grants JWST-GO-2974, HST-GO-16773, and JWST-GO-1794.
This research made use of \textsc{numpy} \citep{Oliphant15}, \textsc{scipy} \citep{Jones01}, \textsc{astropy} \citep{AstropyCollaboration13, AstropyCollaboration18}, \textsc{jupyter} \citep{Kluyver16}, \textsc{matplotlib} \citep{Hunter07}, \textsc{seaborn} \citep{Waskom14}, and \textsc{nautilus} \citep{Lange23}.
\end{acknowledgements}


%
   \bibliographystyle{aa} 
   \bibliography{ajshajib} 

\begin{thebibliography}{30}
\expandafter\ifx\csname natexlab\endcsname\relax\def\natexlab#1{#1}\fi

\bibitem[{{Astropy Collaboration}(2013)}]{AstropyCollaboration13}
{Astropy Collaboration}. 2013, \aap, 558, A33

\bibitem[{{Astropy Collaboration}(2018)}]{AstropyCollaboration18}
{Astropy Collaboration}. 2018, \aj, 156, 123

\bibitem[{B{\"o}ker {et~al.}(2023)B{\"o}ker, Beck, Birkmann, Giardino, Keyes, Kumari, Muzerolle, Rawle, Zeidler, {Abul-Huda}, {Alves de Oliveira}, Arribas, Bechtold, Bhatawdekar, Bonaventura, Bunker, Cameron, Carniani, Charlot, Curti, Espinoza, Ferruit, Franx, Jakobsen, Karakla, {L{\'o}pez-Caniego}, L{\"u}tzgendorf, Maiolino, Manjavacas, Marston, Moseley, Ogle, Perna, {Pe{\~n}a-Guerrero}, Pirzkal, Plesha, Proffitt, Rauscher, Rix, {Rodr{\'i}guez del Pino}, Rustamkulov, Sabbi, Sing, Sirianni, {te Plate}, {\'U}beda, Wahlgren, Wislowski, Wu, \& Willott}]{Boker23}
B{\"o}ker, T., Beck, T.~L., Birkmann, S.~M., {et~al.} 2023, Publications of the Astronomical Society of the Pacific, 135, 038001

\bibitem[{Bushouse {et~al.}(2023)Bushouse, Eisenhamer, Dencheva, Davies, Greenfield, Morrison, Hodge, Simon, Grumm, Droettboom, Slavich, Sosey, Pauly, Miller, Jedrzejewski, Hack, Davis, Crawford, Law, Gordon, Regan, Cara, MacDonald, Bradley, Shanahan, Jamieson, Teodoro, Williams, \& Pena-Guerrero}]{Bushouse23}
Bushouse, H., Eisenhamer, J., Dencheva, N., {et~al.} 2023, JWST Calibration Pipeline

\bibitem[{Cappellari(2017)}]{Cappellari17}
Cappellari, M. 2017, \mnras, 466, 798

\bibitem[{Cappellari(2023)}]{Cappellari23}
Cappellari, M. 2023, \mnras, 526, 3273

\bibitem[{{de Graaff} {et~al.}(2024){de Graaff}, Rix, Carniani, Suess, Charlot, {Curtis-Lake}, Arribas, Baker, Boyett, Bunker, Cameron, Chevallard, Curti, Eisenstein, Franx, Hainline, Hausen, Ji, Johnson, Jones, Maiolino, Maseda, Nelson, Parlanti, Rawle, Robertson, Tacchella, {\"U}bler, Williams, Willmer, \& Willott}]{deGraaff24}
{de Graaff}, A., Rix, H.-W., Carniani, S., {et~al.} 2024, \aap, 684, A87

\bibitem[{D'Eugenio {et~al.}(2024)D'Eugenio, {P{\'e}rez-Gonz{\'a}lez}, Maiolino, Scholtz, Perna, Circosta, {\"U}bler, Arribas, B{\"o}ker, Bunker, Carniani, Charlot, Chevallard, Cresci, {Curtis-Lake}, Jones, Kumari, Lamperti, Looser, Parlanti, Rix, Robertson, Rodr{\'i}guez Del~Pino, Tacchella, Venturi, \& Willott}]{DEugenio24}
D'Eugenio, F., {P{\'e}rez-Gonz{\'a}lez}, P.~G., Maiolino, R., {et~al.} 2024, Nature Astronomy, 8, 1443

\bibitem[{{Euclid Collaboration} {et~al.}(2023){Euclid Collaboration}, Paterson, Schirmer, Copin, Cuillandre, Gillard, Guti{\'e}rrez~Soto, Guzzo, Hoekstra, Kitching, Paltani, Percival, Scodeggio, Stanghellini, Appleton, Laureijs, Mellier, Aghanim, Altieri, Amara, Auricchio, Baldi, Bender, Bodendorf, Bonino, Branchini, Brescia, Brinchmann, Camera, Capobianco, Carbone, Carretero, Castander, Castellano, Cavuoti, Cimatti, Cledassou, Congedo, Conselice, Conversi, Corcione, Courbin, Da~Silva, Degaudenzi, Dinis, Douspis, Dubath, Dupac, Ferriol, Frailis, Franceschi, Fumana, Galeotta, Garilli, Gillis, Giocoli, Grazian, Grupp, Haugan, Holmes, Hornstrup, Hudelot, Jahnke, K{\"u}mmel, Kiessling, Kilbinger, Kohley, Kubik, Kunz, {Kurki-Suonio}, Ligori, Lilje, Lloro, Maiorano, Mansutti, Marggraf, Markovic, Marulli, Massey, Medinaceli, Mei, Meneghetti, Meylan, Moresco, Moscardini, Nakajima, Niemi, Nightingale, Nutma, Padilla, Pasian, Pedersen, Polenta, Poncet, Popa, Raison, Renzi, Rhodes, Riccio, Rix, Romelli, Roncarelli,
  Rossetti, Saglia, Sartoris, Schneider, Secroun, Seidel, Serrano, Sirignano, Sirri, Skottfelt, Stanco, {Tallada-Cresp{\'i}}, Taylor, Tereno, {Toledo-Moreo}, Torradeflot, Tutusaus, Valenziano, Vassallo, Wang, Weller, Zamorani, Zoubian, Andreon, Bardelli, Bozzo, {Colodro-Conde}, Di~Ferdinando, Farina, {Graci{\'a}-Carpio}, Keih{\"a}nen, Lindholm, Maino, Mauri, Scottez, Tenti, Zucca, Akrami, Baccigalupi, Ballardini, Biviano, Borlaff, Burigana, Cabanac, Cappi, Carvalho, Casas, Castignani, Castro, Chambers, Cooray, Coupon, Courtois, Davini, De~Lucia, Desprez, Escartin, Escoffier, Ferrero, Gabarra, {Garcia-Bellido}, George, Giacomini, Gozaliasl, Hildebrandt, Hook, Kajava, Kansal, Kirkpatrick, Legrand, Loureiro, Magliocchetti, Mainetti, Maoli, Marcin, Martinelli, Martinet, Martins, Matthew, Maurin, Metcalf, Monaco, Morgante, Nadathur, Patrizii, Pollack, Porciani, Potter, P{\"o}ntinen, S{\'a}nchez, Sakr, Schneider, Sefusatti, Sereno, Shulevski, Stadel, Steinwagner, \& Valieri}]{EuclidCollaboration23}
{Euclid Collaboration}, Paterson, K., Schirmer, M., {et~al.} 2023, \aap, 674, A172

\bibitem[{Gonneau {et~al.}(2020)Gonneau, Lyubenova, Lan{\c c}on, Trager, Peletier, Arentsen, Chen, Coelho, Dries, {Falc{\'o}n-Barroso}, Prugniel, {S{\'a}nchez-Bl{\'a}zquez}, Vazdekis, \& Verro}]{Gonneau20}
Gonneau, A., Lyubenova, M., Lan{\c c}on, A., {et~al.} 2020, \aap, 634, A133

\bibitem[{Hunter(2007)}]{Hunter07}
Hunter, J.~D. 2007, Computing in Science and Engineering, 9, 90

\bibitem[{Isobe {et~al.}(2023)Isobe, Ouchi, Nakajima, Harikane, Ono, Xu, Zhang, \& Umeda}]{Isobe23}
Isobe, Y., Ouchi, M., Nakajima, K., {et~al.} 2023, \apj, 956, 139

\bibitem[{Jacob {et~al.}(2013)Jacob, Sch{\"o}nberner, \& Steffen}]{Jacob13}
Jacob, R., Sch{\"o}nberner, D., \& Steffen, M. 2013, \aap, 558, A78

\bibitem[{Jones {et~al.}(2001)Jones, Oliphant, Peterson, \& {Others}}]{Jones01}
Jones, E., Oliphant, T., Peterson, P., \& {Others}. 2001, {{SciPy}}: {{Open}} Source Scientific Tools for {{Python}}

\bibitem[{{Jones} {et~al.}(2023){Jones}, {{\'A}lvarez-M{\'a}rquez}, {Sloan}, {Kavanagh}, {Argyriou}, {Law}, {Labiano}, {Patapis}, {Mueller}, {Larson}, {Bright}, {Klaassen}, {Fox}, {Gasman}, {Geers}, {Glauser}, {Guillard}, {Nayak}, {Noriega-Crespo}, {Ressler}, {Sargent}, {Temim}, {Vandenbussche}, \& {Garc{\'\i}a Mar{\'\i}n}}]{Jones23}
{Jones}, O.~C., {{\'A}lvarez-M{\'a}rquez}, J., {Sloan}, G.~C., {et~al.} 2023, \mnras, 523, 2519

\bibitem[{Kluyver {et~al.}(2016)Kluyver, {Ragan-Kelley}, P{\'e}rez, Granger, Bussonnier, Frederic, Kelley, Hamrick, Grout, Corlay, Ivanov, Avila, Abdalla, \& Willing}]{Kluyver16}
Kluyver, T., {Ragan-Kelley}, B., P{\'e}rez, F., {et~al.} 2016, in Positioning and {{Power}} in {{Academic Publishing}}: {{Players}}, {{Agents}} and {{Agendas}}, ed. F.~Loizides \& B.~Schmidt (IOS Press BV, Amsterdam, Netherlands), 87--90

\bibitem[{Knabel {et~al.}(2025)Knabel, Mozumdar, Shajib, Treu, Cappellari, Spiniello, \& Birrer}]{Knabel25}
Knabel, S., Mozumdar, P., Shajib, A.~J., {et~al.} 2025, {{TDCOSMO XIX}}: {{Measuring}} Stellar Velocity Dispersion with Sub-Percent Accuracy for Cosmography

\bibitem[{Lange(2023)}]{Lange23}
Lange, J.~U. 2023, \mnras, 525, 3181

\bibitem[{Modigliani {et~al.}(2010)Modigliani, Goldoni, Royer, Haigron, Guglielmi, Fran{\c c}ois, Horrobin, Bristow, Vernet, Moehler, Kerber, Ballester, Mason, \& Christensen}]{Modigliani10}
Modigliani, A., Goldoni, P., Royer, F., {et~al.} 2010, in Observatory {{Operations}}: {{Strategies}}, {{Processes}}, and {{Systems III}}, Vol. 7737, 773728

\bibitem[{{Newman} {et~al.}(2025){Newman}, {Gu}, {Belli}, {Ellis}, {Gangula}, {Greene}, {Walsh}, {Suyu}, {Ertl}, {Caminha}, {Granata}, {Grillo}, {Schuldt}, {Barone}, {Bird}, {Glazebrook}, {Jafariyazani}, {Kriek}, {Matthews}, {Morishita}, {Nanayakkara}, {Pierel}, {Acebr\textbackslash'on}, {Bergamini}, {Cha}, {Diego}, {Foo}, {Frye}, {Fudamoto}, {Jee}, {Kamieneski}, {Koekemoer}, {Meena}, {Nishida}, {Oguri}, {Rosati}, \& {Zitrin}}]{Newman25}
{Newman}, A.~B., {Gu}, M., {Belli}, S., {et~al.} 2025, arXiv e-prints, arXiv:2503.17478

\bibitem[{{Nidever} {et~al.}(2024){Nidever}, {Gilbert}, {Tollerud}, {Siders}, {Escala}, {Prieto}, {Smith}, {Cunha}, {Debattista}, {Ting}, \& {Kirby}}]{Nidever24}
{Nidever}, D.~L., {Gilbert}, K., {Tollerud}, E., {et~al.} 2024, in IAU Symposium, Vol. 377, Early Disk-Galaxy Formation from JWST to the Milky Way, ed. F.~{Tabatabaei}, B.~{Barbuy}, \& Y.-S. {Ting}, 115--122

\bibitem[{Oliphant(2015)}]{Oliphant15}
Oliphant, T.~E. 2015, Guide to {{NumPy}}, 2nd edn. (USA: CreateSpace Independent Publishing Platform)

\bibitem[{Prochaska {et~al.}(2020)Prochaska, Hennawi, Westfall, Cooke, Wang, Hsyu, Davies, Farina, \& Pelliccia}]{Prochaska20}
Prochaska, J., Hennawi, J., Westfall, K., {et~al.} 2020, JOSS, 5, 2308

\bibitem[{Reid \& Parker(2006)}]{Reid06}
Reid, W.~A. \& Parker, Q.~A. 2006, Monthly Notices of the Royal Astronomical Society, 373, 521

\bibitem[{{Shajib} {et~al.}(2025){Shajib}, {Treu}, {Suyu}, {Law}, {Y{\i}ld{\i}r{\i}m}, {Cappellari}, {Galan}, {Knabel}, {Wang}, {Birrer}, {Courbin}, {Fassnacht}, {Frieman}, {Melo}, {Morishita}, {Mozumdar}, {Sluse}, \& {Stiavelli}}]{Shajib25}
{Shajib}, A.~J., {Treu}, T., {Suyu}, S.~H., {et~al.} 2025, arXiv e-prints, arXiv:2506.21665

\bibitem[{{TDCOSMO Collaboration}(2025)}]{TDCOSMO25}
{TDCOSMO Collaboration}. 2025, arXiv e-prints, arXiv:2506.03023

\bibitem[{{van Hoof}(2018)}]{vanHoof18}
{van Hoof}, P. A.~M. 2018, Galaxies, 6, 63

\bibitem[{Vernet {et~al.}(2011)Vernet, Dekker, D'Odorico, Kaper, Kjaergaard, Hammer, Randich, Zerbi, Groot, Hjorth, Guinouard, Navarro, Adolfse, Albers, Amans, Andersen, Andersen, Binetruy, Bristow, Castillo, Chemla, Christensen, Conconi, Conzelmann, Dam, {de Caprio}, {de Ugarte Postigo}, Delabre, {di Marcantonio}, Downing, Elswijk, Finger, Fischer, Flores, Fran{\c c}ois, Goldoni, Guglielmi, Haigron, Hanenburg, Hendriks, Horrobin, Horville, Jessen, Kerber, Kern, Kiekebusch, Kleszcz, Klougart, Kragt, Larsen, Lizon, Lucuix, Mainieri, Manuputy, Martayan, Mason, Mazzoleni, Michaelsen, Modigliani, Moehler, M{\o}ller, Norup~S{\o}rensen, N{\o}rregaard, P{\'e}roux, Patat, Pena, Pragt, Reinero, Rigal, Riva, Roelfsema, Royer, Sacco, Santin, Schoenmaker, Spano, Sweers, Ter~Horst, Tintori, Tromp, {van Dael}, {van der Vliet}, Venema, Vidali, Vinther, Vola, Winters, Wistisen, Wulterkens, \& Zacchei}]{Vernet11}
Vernet, J., Dekker, H., D'Odorico, S., {et~al.} 2011, \aap, 536, A105

\bibitem[{Waskom {et~al.}(2014)Waskom, Botvinnik, Hobson, Cole, Halchenko, Hoyer, Miles, Augspurger, Yarkoni, Megies, Coelho, Wehner, {cynddl}, Ziegler, {diego0020}, Zaytsev, Hoppe, Seabold, Cloud, Koskinen, Meyer, Qalieh, \& Allan}]{Waskom14}
Waskom, M., Botvinnik, O., Hobson, P., {et~al.} 2014, Seaborn: V0.5.0 ({{November}} 2014)

\bibitem[{Xu {et~al.}(2024)Xu, Ouchi, Yajima, Fukushima, Harikane, Isobe, Nakajima, Nakane, Ono, Umeda, Yanagisawa, \& Zhang}]{Xu24a}
Xu, Y., Ouchi, M., Yajima, H., {et~al.} 2024, \apj, 976, 142

\end{thebibliography}
%

\appendix
\onecolumn

\section{Fitted line lists} \label{app:line_list}

In this appendix, we provide the list of fitted lines within the chosen wavelength ranges in Tables \ref{tab:lines_g140}, \ref{tab:lines_g235}, and \ref{tab:lines_g395} for the G140, G235, and G395 dispersers, respectively.  We first obtained a comprehensive list of nebular \hi\ and \hei\ lines from the Atomic Line List database\footnote{v.300b5: \url{https://linelist.pa.uky.edu/newpage/}} \citep[][]{vanHoof18}. \ajs{We then aggregated \hei\ lines that are close to each other within 1/8th of a pixel size in the corresponding high-resolution spectrum into a single line with an average wavelength weighted by the transition probability (i.e., the Einstein coefficient $g_kA_{ki}$). We adopted 1/8 of the pixel size here, as this is half of the target threshold for wavelength calibration error \citep{Boker23}. Thus, lines that are separated more closely can be sufficiently considered indistinguishable in the spectrum. Not all the line groups were fitted in some modes, as those line groups fall fully or partially outside of the covered wavelength range in the particular modes. The tables in this appendix specify the modes for which each line group was fitted.}

\begin{table*}[!h]
\caption{ \label{tab:lines_g140}
Line list in the fitted wavelength ranges for the G140 disperser.
}
\centering
\renewcommand{\arraystretch}{1.3}
\centering
\begin{tabular}{lcclll}
\hline \hline
Principal Line & $\lambda_{\textrm{min}}$ ($\AA$) & $\lambda_{\textrm{max}}$ ($\AA$) & \hi\ lines ($\AA$) & \hei\ lines ($\AA$) & Mode \\
\hline
Pa$\eta$ & 8980 & 9060 & \begin{tabular}[c]{@{}l@{}}9017.385\end{tabular} & \begin{tabular}[c]{@{}l@{}}8999.448, 9002.207, 9011.625, 9013.688\end{tabular} & FS (070LP) \\
\hline
Pa$\zeta$ & 9190 & 9290 & \begin{tabular}[c]{@{}l@{}}9231.547\end{tabular} & \begin{tabular}[c]{@{}l@{}}9212.861, 9215.757, 9227.763, 9230.392\end{tabular} & FS (070LP) \\
\hline
Pa$\delta$ & 10020 & 10100 & \begin{tabular}[c]{@{}l@{}}10052.128\end{tabular} & \begin{tabular}[c]{@{}l@{}}10025.946, 10030.471, 10033.901\\ 10048.008, 10074.794\end{tabular} & FS (F100LP), IFS \\
\hline
He & 10780 & 10900 & -- & \begin{tabular}[c]{@{}l@{}}10832.057, 10833.273\end{tabular} & FS (F100LP), IFS \\
\hline
Pa$\gamma$ & 10910 & 10990 & \begin{tabular}[c]{@{}l@{}}10941.090\end{tabular} & \begin{tabular}[c]{@{}l@{}}10905.195, 10915.994, 10920.055\\ 10936.607\end{tabular} & FS, IFS \\
\hline
Pa$\beta$ & 12760 & 12885 & \begin{tabular}[c]{@{}l@{}}12821.590\end{tabular} & \begin{tabular}[c]{@{}l@{}}12759.178, 12761.554, 12786.127\\ 12788.433, 12794.005, 12814.100\\ 12816.340, 12846.135, 12849.524\end{tabular} & FS (F100LP), IFS \\
\hline
Br$\zeta$ & 17350 & 17430 & \begin{tabular}[c]{@{}l@{}}17366.850\end{tabular} & \begin{tabular}[c]{@{}l@{}}17340.346, 17356.484, 17358.152\\ 17359.750, 17379.672\end{tabular} & FS (F100LP), IFS \\
\hline
Pa$\alpha$ & 18675 & 18820 & \begin{tabular}[c]{@{}l@{}}18756.130\end{tabular} & \begin{tabular}[c]{@{}l@{}}18690.446, 18702.347, 18730.012\\ 18748.470, 18762.520\end{tabular} & IFS \\
\hline
\end{tabular}
\tablefoot{The $\lambda_{\rm min}$ and $\lambda_{\rm max}$ columns provide the limits of the fitted wavelength ranges. The mode column provides the modes for which the corresponding line group was included in the fit.}
\end{table*}

\begin{table*}
\caption{ \label{tab:lines_g235}
Line list in the fitted wavelength ranges for the G235 disperser.
}
\centering
\renewcommand{\arraystretch}{1.3}
\centering
\begin{tabular}{lcclll}
\hline \hline
Principal Line & $\lambda_{\textrm{min}}$ ($\AA$) & $\lambda_{\textrm{max}}$ ($\AA$) & \hi\ lines ($\AA$) & \hei\ lines ($\AA$) & Mode \\
\hline
Br$\eta$ & 16800 & 16850 & \begin{tabular}[c]{@{}l@{}}16811.111\end{tabular} & \begin{tabular}[c]{@{}l@{}}16801.166, 16802.417, 16804.238\\ 16812.264\end{tabular} & IFS \\
\hline
Br$\zeta$ & 17300 & 17440 & \begin{tabular}[c]{@{}l@{}}17366.850\end{tabular} & \begin{tabular}[c]{@{}l@{}}17332.122, 17334.426, 17340.346\\ 17356.484, 17358.152, 17359.750\\ 17379.672\end{tabular} & FS, IFS \\
\hline
Pa$\alpha$ & 18655 & 18840 & \begin{tabular}[c]{@{}l@{}}18756.130\end{tabular} & \begin{tabular}[c]{@{}l@{}}18690.446, 18702.347, 18730.012\\ 18748.470, 18762.520\end{tabular} & FS, IFS \\
\hline
He & 20555 & 20660 & -- & \begin{tabular}[c]{@{}l@{}}20565.536, 20582.065, 20586.905\\ 20592.793, 20605.478, 20607.440\\ 20622.819\end{tabular} & FS, IFS \\
\hline
Br$\gamma$ & 21575 & 21750 & \begin{tabular}[c]{@{}l@{}}21661.200\end{tabular} & \begin{tabular}[c]{@{}l@{}}21586.002, 21613.713, 21622.911\\ 21633.577, 21647.434, 21652.342\\ 21655.380\end{tabular} & FS, IFS \\
\hline
Br$\beta$ & 26180 & 26350 & \begin{tabular}[c]{@{}l@{}}26258.670\end{tabular} & \begin{tabular}[c]{@{}l@{}}26192.126, 26205.629, 26240.932\\ 26247.940, 26254.427, 26259.100\end{tabular} & FS, IFS \\
\hline
Pf$\eta$ & 30350 & 30480 & \begin{tabular}[c]{@{}l@{}}30392.022\end{tabular} & \begin{tabular}[c]{@{}l@{}}30337.982, 30348.429, 30357.263\\ 30373.932, 30378.344, 30379.607\\ 30399.159, 30408.587\end{tabular} & IFS \\
\hline
\end{tabular}
\tablefoot{The $\lambda_{\rm min}$ and $\lambda_{\rm max}$ columns provide the limits of the fitted wavelength ranges. The mode column provides the modes for which the corresponding line group was included in the fit.}
\end{table*}

\begin{table*}
\caption{ \label{tab:lines_g395}
Line list in the fitted wavelength ranges for the G395 disperser.
}
\centering
\renewcommand{\arraystretch}{1.3}
\centering
\begin{tabular}{lcclll}
\hline \hline
Principal Line & $\lambda_{\textrm{min}}$ ($\AA$) & $\lambda_{\textrm{max}}$ ($\AA$) & \hi\ lines ($\AA$) & \hei\ lines ($\AA$) & Mode \\
\hline
Pf$\eta$ & 30330 & 30520 & \begin{tabular}[c]{@{}l@{}}30392.022\end{tabular} & \begin{tabular}[c]{@{}l@{}}30318.738, 30323.249, 30337.982\\ 30348.429, 30357.263, 30373.932\\ 30378.344, 30379.607, 30399.159\\ 30408.587, 30444.836, 30476.839\end{tabular} & FS, IFS \\
\hline
Pf$\gamma$ & 37350 & 37580 & \begin{tabular}[c]{@{}l@{}}37405.560\\ 37493.933\end{tabular} & \begin{tabular}[c]{@{}l@{}}37328.390, 37344.208, 37381.965\\ 37388.446, 37390.291, 37393.705\\ 37397.995, 37411.662, 37440.355\\ 37442.284, 37450.136, 37473.193\\ 37474.795, 37477.586, 37478.614\\ 37483.145\end{tabular} & FS, IFS \\
\hline
Br$\alpha$ & 40300 & 40730 & \begin{tabular}[c]{@{}l@{}}40522.620\end{tabular} & \begin{tabular}[c]{@{}l@{}}40377.329, 40391.224, 40409.405\\ 40490.196, 40506.090, 40512.200\\ 40544.911, 40563.509, 40574.568\end{tabular} & FS, IFS \\
\hline
He & 42900 & 43100 & -- & \begin{tabular}[c]{@{}l@{}}42954.167, 42959.415\end{tabular} & FS, IFS \\
\hline
Hu$\zeta$ & 43725 & 43900 & \begin{tabular}[c]{@{}l@{}}43764.544\end{tabular} & \begin{tabular}[c]{@{}l@{}}43708.509, 43739.438, 43744.948\\ 43746.575\end{tabular} & FS, IFS \\
\hline
Hu$\delta$ & 51230 & 51450 & \begin{tabular}[c]{@{}l@{}}51286.570\end{tabular} & \begin{tabular}[c]{@{}l@{}}51195.942, 51212.387, 51214.216\\ 51216.767, 51255.944, 51263.452\\ 51265.485, 51267.336, 51271.545\\ 51273.953, 51341.610\end{tabular} & IFS \\
\hline
\end{tabular}
\tablefoot{The $\lambda_{\rm min}$ and $\lambda_{\rm max}$ columns provide the limits of the fitted wavelength ranges. The mode column provides the modes for which the corresponding line group was included in the fit.}
\end{table*}

\end{document}